\documentclass[aps,prl,reprint]{revtex4-2}
\usepackage{amssymb,amsmath,bm,braket,txfonts,graphicx}
\usepackage[bookmarks=true,colorlinks,citecolor=blue,urlcolor=blue]{hyperref}

\begin{document}

\def\papertitle{{Thermal Hall effect and gauge-field picture of magnons in antiferromagnetic skyrmion crystals}}
\def\tum{{Technical University of Munich, TUM School of Natural Sciences, Physics Department, 85748 Garching, Germany}}
\newcommand{\TUM}{\affiliation{\tum}}
\title{\papertitle}
\author{Masataka Kawano} \TUM
\date{\today}

\begin{abstract}
Quasiparticle excitations in material solids often experience a fictitious gauge field, which can be a potential source of intriguing transport phenomena. Here, we show that low-energy excitations in insulating antiferromagnetic skyrmion crystals on the triangular lattice are effectively described by magnons with an SU(3) gauge field. The three-sublattice structure in the antiferromagnetic skyrmion crystals is inherited as three internal degrees of freedom for the magnons, which are coupled to their kinetic motion via the SU(3) gauge field that arises from the topologically nontrivial spin texture in real space. We show that the non-commutativity of the SU(3) gauge field breaks an effective time-reversal symmetry and contributes to a magnon thermal Hall effect. We further demonstrate the emergence of the finite thermal Hall conductivity in the antiferromagnetic skyrmion crystals by the linear spin-wave theory. The possible impact of different gauge structures on a thermal Hall conductivity is also discussed.
\end{abstract}
\maketitle

\emph{\textbf{Introduction.---}}Quasiparticles and gauge fields are fundamental concepts for describing low-energy excitations in material solids. In certain materials, quasiparticles are not simple free particles but those with a fictitious gauge field, leading to anomalous transport phenomena. For example, electrons moving through crystals with noncoplanar spin textures experience a U(1) gauge field, namely a fictitious magnetic field, manifesting as a complex hopping~\cite{wen1989prb,nagaosa1990prl,lee1992prb,matl1998prb,ye1999prl,ohgushi2000prb,chun2000prl,lyanda2001prb,taguchi2001epl,taguchi2001science}. The U(1) gauge field in real space generates a Berry curvature, fictitious magnetic field in momentum space, and contributes to the anomalous Hall effect~\cite{matl1998prb,ye1999prl,ohgushi2000prb,chun2000prl,lyanda2001prb,taguchi2001epl,taguchi2001science}. Magnons, bosonic quasiparticles in magnetic insulators, can also experience a U(1) gauge field that arises from a Dzyaloshinskii-Moriya (DM) interaction or noncoplanar spin textures~\cite{fujimoto2009prl,katsura2010prl,onose2010science,owerre2017prb,laurell2017prl}. Although magnons are charge-neutral and do not feel the Lorentz force, the U(1) gauge field can bend their propagation, leading to the magnon thermal Hall effect~\cite{fujimoto2009prl,katsura2010prl,onose2010science,owerre2017prb,laurell2017prl,matsumoto2011prl,matsumoto2011prb,ideue2012prb,hirschberger2015prl}.

The DM interactions or noncoplanar spin textures typically yield a staggered flux pattern with zero net flux. In this case, the magnon systems adhere to the no-go condition that precludes the magnon thermal Hall effect in edge-shared lattice geometries such as the square and triangular lattices~\cite{katsura2010prl,ideue2012prb}. There, due to the geometrically equivalent cells with opposite fluxes in nearest neighbors, the systems have the effective time-reversal symmetry that leaves the flux pattern unchanged while converting the sign of the thermal Hall conductivity~\cite{katsura2010prl,ideue2012prb}. In contrast, lattices that feature corner-sharing, such as the kagome and pyrochlore lattices, escape this scenario; their geometrically inequivalent neighboring cells allow for the finite thermal Hall conductivity~\cite{katsura2010prl,ideue2012prb}.

Ferromagnetic skyrmion crystals (FM-SkXs), characterized by their topologically nontrivial swirling spin textures, yield a finite net flux, thereby contributing to the anomalous transports even in the edge-shared lattices~\cite{hoogdalem2013prb,kong2013prl,iwasaki2014prb,oh2015prb,roldan2016njp,mook2017prb,kim2019prl,nikolic2020prb,weber2022science,akazawa2022prr}. The magnon thermal Hall effect is observed in the FM-SkX phase of GaV$_{4}$Se$_{8}$~\cite{akazawa2022prr}, where (V$_{4}$Se$_{4}$)$^{5+}$ clusters form the triangular-lattice FM-SkX in the $[111]$ plane~\cite{kezsmarki2015nmat,eugen2015sciad,fujima2017prb,bordacs2017sp}. Recently, the magnon thermal Hall effect has also been observed in the antiferromagnetic skyrmion crystal (AFM-SkX) phase of MnSc$_{2}$S$_{4}$~\cite{takeda2024ncom}, with Mn$^{2+}$ ions forming the triangular-lattice AFM-SkX in the $[111]$ plane~\cite{reil2002jac,fritsch2004prl,gao2017nphys,gao2020nature}. The AFM-SkXs on the triangular lattice consist of three intertwined FM-SkXs that are antiferromagnetically coupled, leading to a three-sublattice structure~\cite{rosales2015prb,diaz2019prl,mukherjee2021scirep,mohylna2022prb,hayami2023arxiv}. The net flux in the AFM-SkXs, however, is naively expected to be zero because of their antiferromagnetic coupling. To explain the origin of the magnon thermal Hall effect in the AFM-SkXs, the authors in Ref.~\cite{takeda2024ncom} introduce the concept of an SU(3) gauge field of magnons. However, how the spin textures in the AFM-SkXs are translated into the SU(3) gauge field has not yet been clarified well.

In this Letter, we bridge this gap by constructing the effective field theory of magnons in the AFM-SkXs from a spin model on the triangular lattice. We show that the three sublattices in the AFM-SkXs introduce the three internal degrees of freedom for the magnons, which are coupled to their kinetic motion via the SU(3) gauge field originating from the combined effect of the spin textures of three FM-SkXs and the local antiferromagnetic coupling between them. We find that the commutation relation of the SU(3) gauge field generates a uniformly distributed flux, which is responsible for the magnon thermal Hall effect. Using the linear spin-wave theory, we also demonstrate that the thermal Hall conductivity remains very small in the helical phase but becomes substantially large in the AFM-SkX phase. The possible impact of the different gauge structures in FM-SkXs and AFM-SkXs on the thermal Hall conductivity is also discussed.

\begin{figure}[t]
    \includegraphics[width=85mm]{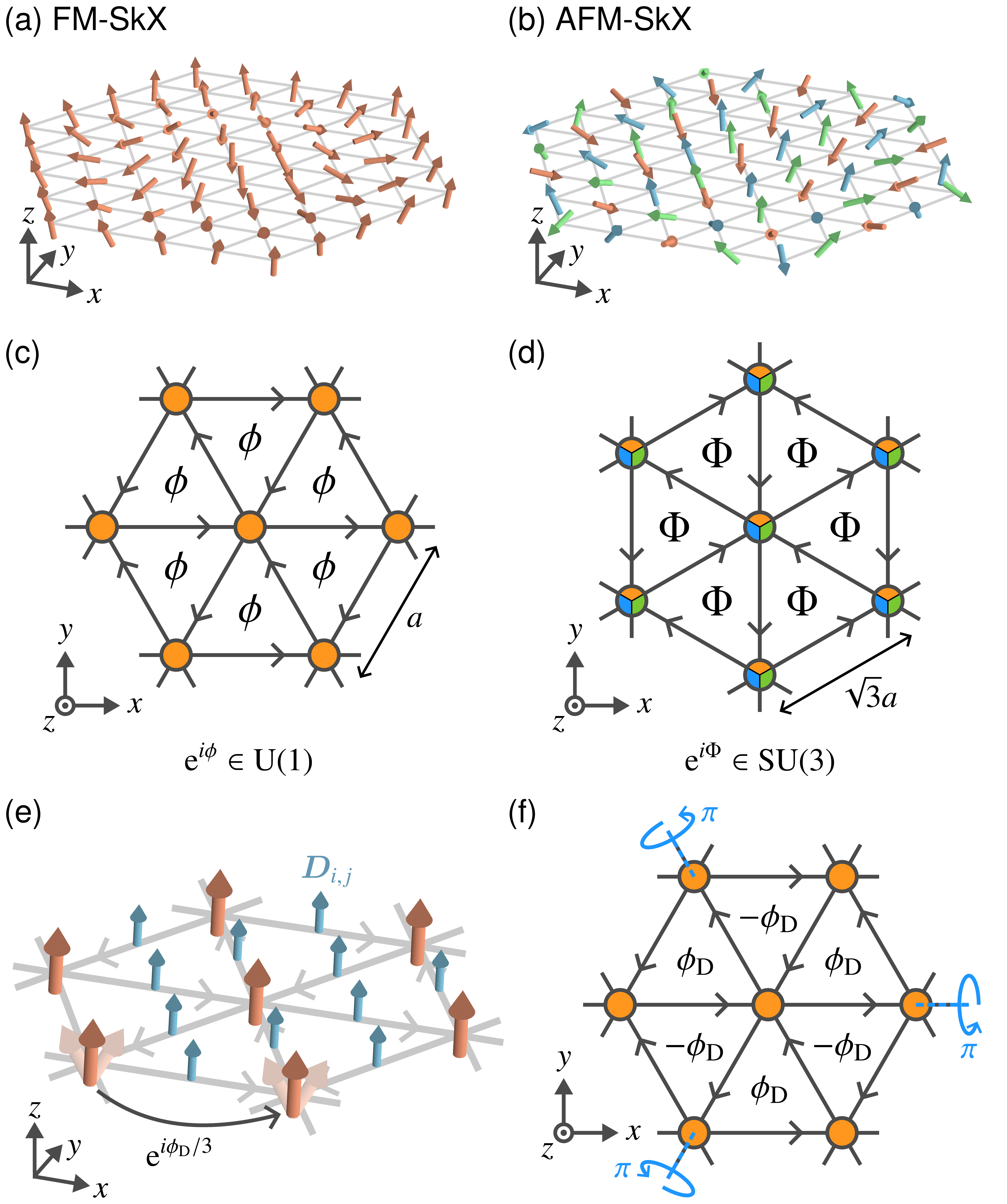}
    \caption{\textbf{Skyrmion crystals and gauge fields of magnons.} (a), (b) Schematic illustration of the real-space spin configuration of the (a) FM-SkXs and (b) AFM-SkXs on the triangular lattice. The AFM-SkXs consist of three intertwined FM-SkXs as shown in different colors, giving rise to the three-sublattice structure. (c), (d) Schematic illustration of the uniform flux in the low-energy magnon systems for the (c) FM-SkXs and (d) AFM-SkXs. The clockwise (counterclockwise) arrows indicate the negative (positive) direction of the flux. The three different colors on each site in (d) represent the three internal degrees of freedom for the magnons. The uniform flux $\Phi$ originates from the commutation relation of an SU(3) gauge field. (e) Schematic illustration of the triangular-lattice ferromagnet with the DM interaction $\bm{D}_{i,j}\parallel\bm{e}^{Z}$. Gray arrows indicate the order of spins in the DM term. (f) Staggered flux pattern generated by the DM interaction. The system has the $\pi$-rotation symmetry along any straight line in the triangular lattice, which prohibits the finite thermal Hall conductivity.}
    \label{fig:skx}
\end{figure}

\emph{\textbf{Effective field theory of magnons in FM-SkXs.---}}Before we explore the effective field theory of magnons in the AFM-SkXs, it is instructive first to briefly review that in the FM-SkXs~\cite{hoogdalem2013prb,kong2013prl,iwasaki2014prb,oh2015prb,roldan2016njp,mook2017prb,kim2019prl,nikolic2020prb,weber2022science,akazawa2022prr}. We start from a spin Hamiltonian on the triangular lattice,
\begin{align}
    \hat{\mathcal{H}}=\sum_{\braket{i,j}}\left[J\hat{\bm{S}}_{i}\cdot\hat{\bm{S}}_{j}+\bm{D}_{i,j}\cdot(\hat{\bm{S}}_{i}\times\hat{\bm{S}}_{j})\right]-g\mu_{\mathrm{B}}B\sum_{i}\hat{S}_{i}^{z},
    \label{eq:H}
\end{align}
where $\braket{i,j}$ denotes the pair of nearest-neighbor $i$ and $j$ sites, $\hat{\bm{S}}_{i}=(\hat{S}_{i}^{x},\hat{S}_{i}^{y},\hat{S}_{i}^{z})$ is the spin-$S$ operator at site $i$, $J$ is the Heisenberg exchange coupling constant, $\bm{D}_{i,j}=D\bm{e}^{Z}\times\bm{e}_{i,j}$ is the DM vector, $\bm{e}^{Z}$ is the unit vector pointing in the $z$-direction in spin space, $\bm{e}_{i,j}$ is the unit vector pointing from site $i$ to site $j$, and $B$ is the magnetic field in $z$-direction with g-factor $g$ and Bohr magneton $\mu_{\mathrm{B}}$. Various spin configurations can be realized in noncentrosymmetric magnets by the competition between exchange parameters, anisotropy, and magnetic field~\cite{bogdanov1989jetp,bogdanov2002prb}. For sufficiently large $D/|J|$ and $g\mu_{\mathrm{B}}B/|J|$, the FM-SkX (AFM-SkX) phase is realized in the ground state of Eq.~(\ref{eq:H}) for $J<0$ ($J>0$)~\cite{rosales2015prb,diaz2019prl,diaz2020prr}. The schematic illustrations of the FM-SkX and AFM-SkX are shown in Fig.~\ref{fig:skx}(a) and Fig.~\ref{fig:skx}(b).

Given the slow spatial variation of spin orientations in the FM-SkXs (see Fig.~\ref{fig:skx}(a)), the low-energy theory is expected to be described by a continuously varying spin-density operator $\hat{\bm{s}}(\bm{r})$. Adopting this assumption, the effective field theory is derived by substituting $\hat{\bm{S}}_{i}$ with $v\hat{\bm{s}}(\bm{r})$ and $\sum_{i}$ with $(1/v)\int \mathrm{d}^{2}\bm{r}$, where $v=\sqrt{3}a^{2}/2$ is the volume per site with lattice constant $a$. Applying these substitutions to Eq.~(\ref{eq:H}) leads to the effective Hamiltonian
\begin{align}
    \hat{\mathcal{H}}_{\mathrm{eff}}^{\mathrm{FM}}=\int\mathrm{d}^{2}\bm{r}\left[\hat{\bm{s}}^{T}(\bm{r})K(\bm{r})\hat{\bm{s}}(\bm{r})-g\mu_{\mathrm{B}}B\hat{s}^{z}(\bm{r})\right],
    \label{eq:Heff_FM_s}
\end{align}
where $K(\bm{r})$ is the $3\times3$ matrix defined as
\begin{align}
    K(\bm{r})=3Jv\left(1+\frac{a^{2}}{4}\nabla^{2}\right)I+\frac{3}{2}Dva
    \begin{pmatrix}
        0 & 0 & -\partial_{x}\\
        0 & 0 & -\partial_{y}\\
        \partial_{x} & \partial_{y} & 0
    \end{pmatrix}
    ,
\end{align}
with the $3\times3$ identity matrix $I$. Here, we drop higher-order derivative terms, which are irrelevant at low energies. To describe the magnon excitations, we employ the Holstein-Primakoff transformation~\cite{holstein19401940}
\begin{align}
    \hat{\bm{s}}(\bm{r})&\simeq\sqrt{\frac{S}{v}}\left(\hat{b}(\bm{r})\bm{e}^{-}(\bm{r})+\mathrm{H.c.}\right)+\left(\frac{S}{v}-\hat{b}^{\dagger}(\bm{r})\hat{b}(\bm{r})\right)\bm{m}(\bm{r}),
    \label{eq:HPtr}
\end{align}
where $\hat{b}(\bm{r})$ ($\hat{b}^{\dagger}(\bm{r})$) is the magnon annihilation (creation) operator at $\bm{r}$, $\bm{m}(\bm{r})=(m^{x}(\bm{r}),m^{y}(\bm{r}),m^{z}(\bm{r}))$ is the unit vector pointing in the direction of the spin in the classical ground state, and $\bm{e}^{\pm}(\bm{r})=(\bm{e}^{x}(\bm{r})\pm i\bm{e}^{y}(\bm{r}))/\sqrt{2}$ with the unit vectors, $\bm{e}^{x}(\bm{r})$ and $\bm{e}^{y}(\bm{r})$, satisfying $\bm{e}^{x}(\bm{r})\times\bm{e}^{y}(\bm{r})=\bm{m}(\bm{r})$. The three unit vectors, $\bm{e}^{x}(\bm{r})$, $\bm{e}^{y}(\bm{r})$, and $\bm{m}(\bm{r})$, form a local orthonormal basis. We neglect constant terms, terms quadratic in the derivative of $\bm{e}^{\pm}(\bm{r})$, and magnon-magnon interaction terms, which are only relevant at high energies~\cite{zhitomirsky2013rmp,mook2020prr}. From Eqs.~(\ref{eq:Heff_FM_s}) and~(\ref{eq:HPtr}), we obtain the effective magnon Hamiltonian as
\begin{align}
    \hat{\mathcal{H}}_{\mathrm{eff}}^{\mathrm{FM}}\simeq\int\mathrm{d}^{2}\bm{r}\ \hat{b}^{\dagger}(\bm{r})\left[\frac{3}{2}JSa^{2}\left(\bm{\nabla}-i\bm{A}(\bm{r})\right)^{2}-\hbar A_{0}(\bm{r})\right]\hat{b}(\bm{r}),
    \label{eq:Heff_FM_b}
\end{align}
with U(1) gauge fields $A_{0}(\bm{r})=-g\mu_{\mathrm{B}}Bm^{z}(\bm{r})/\hbar$, $A_{x}(\bm{r})=i\bm{e}^{+}(\bm{r})\cdot\partial_{x}\bm{e}^{-}(\bm{r})+(D/Ja)m^{y}(\bm{r})$, and $A_{y}(\bm{r})=i\bm{e}^{+}(\bm{r})\cdot\partial_{y}\bm{e}^{-}(\bm{r})-(D/Ja)m^{x}(\bm{r})$. The Mermin-Ho relation~\cite{mermin1976prl} expresses the associated fictitious magnetic field, $B(\bm{r})=\partial_{x}A_{y}(\bm{r})-\partial_{y}A_{x}(\bm{r})$, in terms of $\bm{m}(\bm{r})$ as
\begin{align}
    B(\bm{r})=\bm{m}(\bm{r})\cdot\left(\partial_{x}\bm{m}(\bm{r})\times\partial_{y}\bm{m}(\bm{r})\right)-\frac{D}{Ja}\sum_{\mu=x,y}\partial_{\mu}m^{\mu}(\bm{r}),
\end{align}
which is the gauge-invariant, physical field coupled to the magnons~\cite{hoogdalem2013prb,kong2013prl,iwasaki2014prb,oh2015prb,roldan2016njp,mook2017prb,kim2019prl,nikolic2020prb,weber2022science,akazawa2022prr}. Its uniform component is proportional to the skyrmion density $\rho_{\mathrm{FM-SkX}}=(1/4\pi V)\int\mathrm{d}^{2}\bm{r}\ \bm{m}(\bm{r})\cdot(\partial_{x}\bm{m}(\bm{r})\times\partial_{y}\bm{m}(\bm{r}))$ and takes finite value in the FM-SkX, where $V$ is the system's volume. In terms of the original lattice, the uniform component gives rise to uniformly distributed flux $\phi=2\pi v\rho_{\mathrm{FM-SkX}}$ as illustrated in Fig.~\ref{fig:skx}(c). Therefore, the situation is analogous to that of charged particles moving in a uniform magnetic field, and the magnon thermal Hall effect emerges in a manner similar to the well-known electronic Hall effect~\cite{hoogdalem2013prb,kong2013prl,iwasaki2014prb,oh2015prb,roldan2016njp,mook2017prb,kim2019prl,nikolic2020prb,weber2022science,akazawa2022prr}.

The emergence of the uniform flux is particularly important since otherwise the no-go condition precludes a finite thermal Hall conductivity~\cite{katsura2010prl,ideue2012prb}. We briefly review this fact by considering the triangular-lattice ferromagnet with the DM interaction shown in Fig.~\ref{fig:skx}(e). The DM interaction leads to the complex hopping of magnons $\mathrm{e}^{i\phi_{\mathrm{D}}/3}\hat{b}_{i}^{\dagger}\hat{b}_{j}$ with $\phi_{\mathrm{D}}/3=-\arctan(D/J)$, resulting in the staggered flux pattern shown in Fig.~\ref{fig:skx}(f). The flux pattern is preserved under a $\pi$-rotation along any straight line in the triangular lattice. This rotation, however, converts the sign of the thermal Hall conductivity, indicating that the DM-induced flux does not contribute to the magnon thermal Hall effect.

\begin{figure}[t]
    \includegraphics[width=85mm]{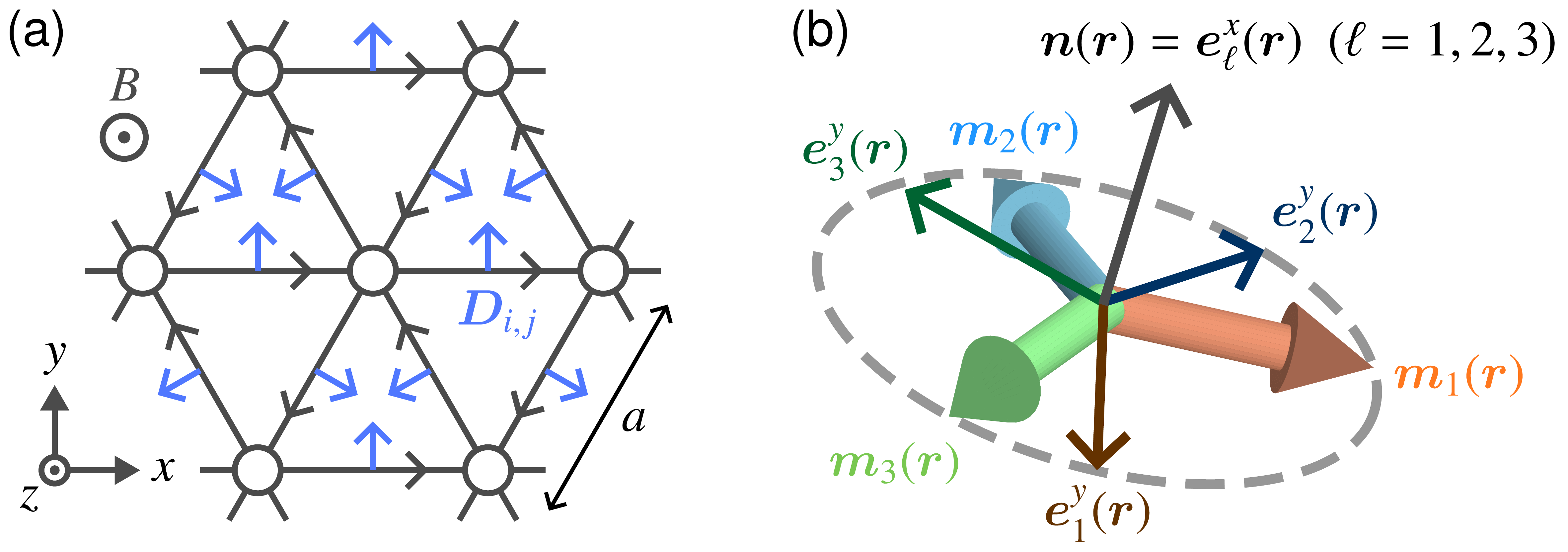}
    \caption{\textbf{Model and orthogonal basis.} (a) Schematic illustration of the spin model on the triangular lattice. The DM vectors $\bm{D}_{i,j}$ are marked by blue arrows and the corresponding bond directions $\bm{e}_{i,j}$ by black arrows. (b) Three local orthonormal bases in the AFM-SkXs. The three unit vectors $\bm{m}_{\ell}(\bm{r})$ ($\ell=1,2,3$) are approximately on the same plane. We take the unit vector $\bm{n}(\bm{r})=\bm{e}_{\ell}^{x}(\bm{r})$ to be orthogonal to this plane.}
    \label{fig:model}
\end{figure}

\emph{\textbf{Effective field theory of magnons in AFM-SkXs.---}}We turn to the effective field theory of magnons in the AFM-SkXs. A key observation is that, since the AFM-SkXs consist of three intertwined FM-SkXs, we need to introduce three continuously varying spin-density operators $\hat{\bm{s}}_{\ell}(\bm{r})$ with sublattice index $\ell=1,2,3$ to describe the low-energy excitations~\cite{takeda2024ncom}. The spin operator $\hat{\bm{S}}_{i}$ is substituted with $v'\hat{\bm{s}}_{\ell}(\bm{r})$ when site $i$ belongs to the sublattice $\ell$, and $\sum_{i}$ is substituted with $(1/v')\int \mathrm{d}^{2}\bm{r}\sum_{\ell}$, where $v'=3v$. By these substitutions, we obtain the effective Hamiltonian from Eq.~(\ref{eq:H}) as
\begin{align}
    \hat{\mathcal{H}}_{\mathrm{eff}}^{\mathrm{AFM}}=\hspace{-3pt}\int\mathrm{d}^{2}\bm{r}\hspace{-1pt}\sum_{\ell}\left[\frac{3Jv'}{2}\hspace{-4pt}\sum_{\ell'(\neq\ell)}\hat{\bm{s}}_{\ell}^{T}(\bm{r})K(\bm{r})\hat{\bm{s}}_{\ell'}(\bm{r})-g\mu_{\mathrm{B}}B\hat{s}_{\ell}^{z}(\bm{r})\right].
    \label{eq:H_eff_AFM_s}
\end{align}
As in the case of the FM-SkXs, we define a local orthonormal basis, $\bm{e}_{\ell}^{x}(\bm{r})$, $\bm{e}_{\ell}^{y}(\bm{r})$, and $\bm{m}_{\ell}(\bm{r})$, with sublattice index $\ell$ as shown in Fig.~\ref{fig:model}(b). The unit vector $\bm{m}_{\ell}(\bm{r})$ denotes the direction of the spin at $\bm{r}$ with index $\ell$. To incorporate the local $120^{\circ}$ order in the AFM-SkXs, we impose the local constraint $\sum_{\ell}\bm{m}_{\ell}(\bm{r})\simeq 3m_{\mathrm{u}}^{z}\bm{e}^{Z}$ with a small uniform magnetization $m_{\mathrm{u}}^{z}\ll1$, indicating that the three unit vectors $\bm{m}_{\ell}(\bm{r})$ are approximately on the same plane. We introduce the unit vector $\bm{n}(\bm{r})$ that is orthogonal to this plane as $\bm{n}(\bm{r})=(2/\sqrt{3})\bm{m}_{1}(\bm{r})\times\bm{m}_{2}(\bm{r})$, and take $\bm{e}_{\ell}^{x}(\bm{r})=\bm{n}(\bm{r})$ for all $\ell=1,2,3$, which simplifies the following calculations. Applying the Holstein-Primaloff transformation (\ref{eq:HPtr}) with index $\ell$ and unit volume $v'$ to Eq.~(\ref{eq:H_eff_AFM_s}) leads to the effective magnon Hamiltonian
\begin{align}
    \hat{\mathcal{H}}_{\mathrm{eff}}^{\mathrm{AFM}}\simeq\frac{1}{2}\int \mathrm{d}^{2}\bm{r}\sum_{\ell,\ell'}
    \begin{pmatrix}
        \hat{b}_{\ell}^{\dagger}(\bm{r}) & \hat{b}_{\ell}(\bm{r})
    \end{pmatrix}
    \tilde{H}_{\ell,\ell'}(\bm{r})
    \begin{pmatrix}
        \hat{b}_{\ell'}(\bm{r})\\
        \hat{b}_{\ell'}^{\dagger}(\bm{r})
    \end{pmatrix}
    ,
    \label{eq:Heff_AFM_b}
\end{align}
where $\hat{b}_{\ell}(\bm{r})$ ($\hat{b}_{\ell}^{\dagger}(\bm{r})$) is the magnon annihilation (creation) operator with sublattice index $\ell$ and $\tilde{H}_{\ell,\ell'}(\bm{r})$ is the $2\times2$ matrix defined as $\tilde{H}_{\ell,\ell}(\bm{r})=[3JS+g\mu_{\mathrm{B}}Bm_{\ell}^{z}(\bm{r})]\tau^{0}$ for $\ell=\ell'$ and
\begin{align}
    \tilde{H}_{\ell,\ell'}(\bm{r})&=\frac{3JS}{4}\left(1+\frac{a^{2}}{4}\nabla^{2}\right)(\tau^{0}+3\tau^{x}) \nonumber \\
    &+\frac{3JSa^{2}}{8}\left[\bm{A}_{\ell}(\bm{r})(\tau^{y}-i\tau^{z})-\bm{A}_{\ell'}(\bm{r})(\tau^{y}+i\tau^{z})\right]\cdot\bm{\nabla} \nonumber \\
    &+\frac{3JSa^{2}}{8}\left(\sum_{\ell''}\epsilon_{\ell\ell'\ell''}\right)\bm{\xi}(\bm{r})(\tau^{0}-\tau^{x})\cdot\bm{\nabla},
\end{align}
for $\ell\neq\ell'$ with unit and Pauli matrices $\tau^{\nu}$ ($\nu=0,x,y,z$) and Levi-Civita symbol $\epsilon_{\ell\ell'\ell''}$. Here, we introduce four vector fields as $A_{\ell,x}(\bm{r})=i\bm{e}_{\ell}^{+}(\bm{r})\cdot\partial_{x}\bm{e}_{\ell}^{-}(\bm{r})+(2D/Ja)m_{\ell}^{y}(\bm{r})$, $A_{\ell,y}(\bm{r})=i\bm{e}_{\ell}^{+}(\bm{r})\cdot\partial_{y}\bm{e}_{\ell}^{-}(\bm{r})-(2D/Ja)m_{\ell}^{x}(\bm{r})$, and $\xi_{\mu}(\bm{r})=(1/3)[\bm{m}_{1}(\bm{r})\cdot\partial_{\mu}\bm{m}_{2}(\bm{r})+\bm{m}_{2}(\bm{r})\cdot\partial_{\mu}\bm{m}_{3}(\bm{r})+\bm{m}_{3}(\bm{r})\cdot\partial_{\mu}\bm{m}_{1}(\bm{r})]$ ($\mu=x,y$). In particular, $\bm{A}_{\ell}(\bm{r})$ can be interpreted as the U(1) gauge field on the sublattice $\ell$.

The effective magnon Hamiltonian~(\ref{eq:Heff_AFM_b}) is complicated and conceals its gauge structure. To elucidate the hidden gauge structure, we introduce new bosonic operators $\hat{\gamma}_{m}(\bm{r})$ ($m=1,2,3$) as the linear combination of $\hat{b}_{\ell}(\bm{r})$ and $\hat{b}_{\ell}^{\dagger}(\bm{r})$ such that the resulting effective Hamiltonian has the following two properties; (i) the conventional $120^{\circ}$ order, namely spatially uniform $\bm{m}_{\ell}(\bm{r})$, gives rise to three decoupled magnons with well-known linear dispersions and (ii) the equation of motion for $\hat{\gamma}_{m}(\bm{r})$ is relativistic, which is expected in antiferromagnets. We find the Bogoliubov transformation that satisfies these properties as
\begin{align}
    \hat{\gamma}_{1}(\bm{r})&=\frac{1}{\sqrt{3}}\sum_{\ell}\left(\cosh\chi\hat{b}_{\ell}(\bm{r})+\sinh\chi\hat{b}_{\ell}^{\dagger}(\bm{r})\right),\\
    \hat{\gamma}_{2}(\bm{r})&=i\sqrt{\frac{2}{3}}\sum_{\ell}\cos\theta_{\ell}\left(\cosh\chi\hat{b}_{\ell}(\bm{r})+\sinh\chi\hat{b}_{\ell}^{\dagger}(\bm{r})\right),\\
    \hat{\gamma}_{3}(\bm{r})&=i\sqrt{\frac{2}{3}}\sum_{\ell}\sin\theta_{\ell}\left(\cosh\chi\hat{b}_{\ell}(\bm{r})+\sinh\chi\hat{b}_{\ell}^{\dagger}(\bm{r})\right),
\end{align}
with $\theta_{\ell}=2\pi(\ell-1)/3$ and $\chi=(1/2)\mathrm{arctanh}(1/3)$. The effective Hamiltonian (\ref{eq:Heff_AFM_b}) is transformed as
\begin{align}
    \hat{\mathcal{H}}_{\mathrm{eff}}^{\mathrm{AFM}}&\simeq\frac{1}{2}\int \mathrm{d}^{2}\bm{r}\ \hat{\Psi}^{\dagger}(\bm{r})H(\bm{r})\hat{\Psi}(\bm{r}),
    \label{eq:Heff_AFM_gamma}
\end{align}
where $\hat{\Psi}(\bm{r})=(\hat{\gamma}_{1}(\bm{r}),\hat{\gamma}_{1}^{\dagger}(\bm{r}),\hat{\gamma}_{2}(\bm{r}),\hat{\gamma}_{2}^{\dagger}(\bm{r}),\hat{\gamma}_{3}(\bm{r}),\hat{\gamma}_{3}^{\dagger}(\bm{r}))^{T}$ and
\begin{align}
    H(\bm{r})&=\frac{3\sqrt{2}}{8}JSa^{2}\left[\tilde{I}\otimes\tau^{x}\nabla^{2}-i\bm{G}(\bm{r})\cdot\bm{\nabla}\right] \nonumber \\
    &\hspace{20pt}+\frac{9\sqrt{2}}{4}JSI\otimes(\tau^{0}+\tau^{x})+g\mu_{\mathrm{B}}BM,
    \label{eq:Hr}
\end{align}
with $\tilde{I}=\mathrm{diag}(2,1,1)$. The $6\times6$ matrices $\bm{G}(\bm{r})=(G_{x}(\bm{r}),G_{y}(\bm{r}))$ and $M$ are defined as
\begin{align}
    \bm{G}(\bm{r})&=\bm{A}_{\mathrm{c}}(\bm{r})\left[\lambda_{2}\otimes(\tau^{0}+3\tau^{x})+\frac{1}{\sqrt{2}}(\lambda_{3}-\sqrt{3}\lambda_{8})\otimes\tau^{z}\right]\nonumber\\
    &\quad+\bm{A}_{\mathrm{s}}(\bm{r})\left[\lambda_{5}\otimes(\tau^{0}+3\tau^{x})+\sqrt{2}\lambda_{6}\otimes\tau^{z}\right]\nonumber\\
    &\quad-\sqrt{3}\bm{\xi}(\bm{r})\lambda_{7}\otimes(\tau^{0}+\tau^{x}),\\
    M&=\frac{\sqrt{2}}{12}m_{\mathrm{u}}^{z}\left[I\otimes(9\tau^{0}+\tau^{x})-\sqrt{3}(\sqrt{3}\lambda_{3}+\lambda_{8})\otimes\tau^{x}\right],
\end{align}
where $\bm{A}_{\mathrm{c}}(\bm{r})=(1/3)\sum_{\ell}\bm{A}_{\ell}(\bm{r})\cos\theta_{\ell}$, $\bm{A}_{\mathrm{s}}(\bm{r})=(1/3)\sum_{\ell}\bm{A}_{\ell}(\bm{r})\sin\theta_{\ell}$, and $\lambda_{\alpha}$ ($\alpha=1,2,\ldots,8$) are the Gell-Mann matrices. Here, we ignore terms proportional to $\sum_{\ell}\bm{A}_{\ell}(\bm{r})\simeq0$, $m_{\mathrm{c}}^{z}(\bm{r})=(1/3)\sum_{\ell}m_{\ell}^{z}(\bm{r})\cos\theta_{\ell}$, and $m_{\mathrm{s}}^{z}(\bm{r})=(1/3)\sum_{\ell}m_{\ell}^{z}(\bm{r})\sin\theta_{\ell}$. The latter two values vanish upon spatial averaging and only give higher-order contributions to the equation of motion for $\hat{\gamma}_{m}(\bm{r})$.

The gauge structure becomes more apparent upon deriving the equation of motion for $\hat{\bm{\gamma}}(\bm{r})=(\hat{\gamma}_{1}(\bm{r}),\hat{\gamma}_{2}(\bm{r}),\hat{\gamma}_{3}(\bm{r}))^{T}$. From the effective Hamiltonian (\ref{eq:Heff_AFM_gamma}), we obtain the equation of motion for $\hat{\Psi}(\bm{r},t)$ as $i\hbar\partial_{t}\hat{\Psi}(\bm{r},t)=(I\otimes\tau^{z})H(\bm{r})\hat{\Psi}(\bm{r},t)$. Then the equation of motion for $\hat{\bm{\gamma}}(\bm{r},t)$ is derived as~\cite{supp}
\begin{align}
    \hbar^{2}\frac{\partial^{2}}{\partial t^{2}}\hat{\bm{\gamma}}(\bm{r},t)=\left[\frac{27}{8}(JSa)^{2}\tilde{I}\left(\bm{\nabla}I-i\bm{T}(\bm{r})\right)^{2}+\Xi\right]\hat{\bm{\gamma}}(\bm{r},t),
    \label{eq:eom_gamma}
\end{align}
where the $3\times3$ matrices $\bm{T}(\bm{r})=(T_{x}(\bm{r}),T_{y}(\bm{r}))$ and $\Xi$ are defined as
\begin{align}
    \tilde{I}\bm{T}(\bm{r})&=\bm{A}_{\mathrm{c}}(\bm{r})\left[\lambda_{2}-\frac{\sqrt{2}}{4}(\lambda_{3}-\sqrt{3}\lambda_{8})\right]+\bm{A}_{\mathrm{s}}(\bm{r})\left(\lambda_{5}-\frac{\lambda_{6}}{\sqrt{2}}\right)
    \label{eq:T}\\
    \Xi&=\frac{3}{4}JSg\mu_{\mathrm{B}}Bm_{\mathrm{u}}^{z}\left[8I+\sqrt{3}(\sqrt{3}\lambda_{3}+\lambda_{8})\right].
\end{align}
Since $T_{\mu}(\bm{r})$ is Hermitian and traceless, it belongs to the Lie algebra of the SU(3) group and can be interpreted as the SU(3) gauge field. The diagonal matrix $\Xi$ describes the potential term from the coupling between the magnetic field and uniform magnetization. For the conventional $120^{\circ}$ order, namely for spatially-independent $\bm{m}_{\ell}(\bm{r})$ with $m_{\mathrm{u}}^{z}=0$, we have $\bm{T}(\bm{r})=\bm{0}$ and $\Xi=0$, reproducing three independent magnons with linear dispersions $\varepsilon_{1}(\bm{k})\simeq\sqrt{2}(3/2)^{3/2}JSa|\bm{k}|$ and $\varepsilon_{2}(\bm{k})=\varepsilon_{3}(\bm{k})\simeq(3/2)^{3/2}JSa|\bm{k}|$~\cite{jolicoeur1989prb}.

The SU(3) gauge field $\bm{T}(\bm{r})$ generates a physical field, called field strength, defined as $F_{xy}(\bm{r})=\partial_{x}T_{y}(\bm{r})-\partial_{y}T_{x}(\bm{r})-i[T_{x}(\bm{r}),T_{y}(\bm{r})]$, which is the counterpart of the fictitious magnetic field in the U(1) gauge field. Since $\tilde{I}F_{xy}(\bm{r})$ belongs to the Lie algebra of the SU(3) group, it can be expanded as $\tilde{I}F_{xy}(\bm{r})=\sum_{\alpha=1}^{8}\tilde{F}_{xy}^{(\alpha)}(\bm{r})\lambda_{\alpha}$. The field strength is characterized by the eight real values $\tilde{F}_{xy}^{(\alpha)}(\bm{r})$, whose nonzero elements are calculated from Eq.~(\ref{eq:T}) as
\begin{align}
    &\tilde{F}_{xy}^{(2)}(\bm{r})=B_{\mathrm{c}}(\bm{r}), \hspace{5pt} \tilde{F}_{xy}^{(3)}(\bm{r})=-\frac{\sqrt{2}}{4}B_{\mathrm{c}}(\bm{r}), \hspace{5pt} \tilde{F}_{xy}^{(8)}(\bm{r})=\frac{\sqrt{6}}{4}B_{\mathrm{c}}(\bm{r})
    \label{eq:F2,F3,F8}\\
    &\tilde{F}_{xy}^{(5)}(\bm{r})=B_{\mathrm{s}}(\bm{r}), \hspace{5pt} \tilde{F}_{xy}^{(6)}(\bm{r})=-\frac{B_{\mathrm{s}}(\bm{r})}{\sqrt{2}}
    \label{eq:F5,F6}\\
    &\tilde{F}_{xy}^{(7)}(\bm{r})=-\frac{1}{8}\bm{n}(\bm{r})\cdot(\partial_{x}\bm{n}(\bm{r})\times\partial_{y}\bm{n}(\bm{r}))\nonumber\\
    &\hspace{37.5pt}-\frac{1}{3}\frac{D}{Ja}\left([\bm{n}(\bm{r})\times\partial_{x}\bm{n}(\bm{r})]^{x}+[\bm{n}(\bm{r})\times\partial_{y}\bm{n}(\bm{r})]^{y}\right)\nonumber\\
    &\hspace{37.5pt}-\frac{\sqrt{3}}{12}\left(\frac{2D}{Ja}\right)^{2}n^{z}(\bm{r}),
    \label{eq:F7}
\end{align}
where $B_{\mathrm{c}}(\bm{r})=(1/3)\sum_{\ell}B_{\ell}(\bm{r})\cos\theta_{\ell}$, $B_{\mathrm{s}}(\bm{r})=(1/3)\sum_{\ell}B_{\ell}(\bm{r})\sin\theta_{\ell}$, and $B_{\ell}(\bm{r})=\bm{m}_{\ell}(\bm{r})\cdot(\partial_{x}\bm{m}_{\ell}(\bm{r})\times\partial_{y}\bm{m}_{\ell}(\bm{r}))-(2D/Ja)\sum_{\mu=x,y}\partial_{\mu}m_{\ell}^{\mu}(\bm{r})$ is the fictitious magnetic field on the sublattice $\ell$. In particular, $\tilde{F}_{xy}^{(7)}(\bm{r})$ comes from the commutation relation $[T_{x}(\bm{r}),T_{y}(\bm{r})]$ and is determined by the scalar chirality of the vector field $\bm{n}(\bm{r})$ and DM interaction.

In analogy to the case of the FM-SkXs, we focus on the uniform elements of the field strength by replacing $\tilde{F}_{xy}^{(\alpha)}(\bm{r})$ with $\bar{F}_{xy}^{(\alpha)}=(1/V)\int \mathrm{d}^{2}\bm{r}\ \tilde{F}_{xy}^{(\alpha)}(\bm{r})$. We assume $\bar{B}_{1}=\bar{B}_{2}=\bar{B}_{3}$ since three FM-SkXs in the AFM-SkXs typically have the same skyrmion density~\cite{rosales2015prb,diaz2019prl,mukherjee2021scirep,mohylna2022prb}. Within this approximation, the terms proportional to $B_{\mathrm{c}}(\bm{r})$ and $B_{\mathrm{s}}(\bm{r})$ are reduced to zero, whereas $\bar{F}_{xy}^{(7)}$ takes the finite value. The first term in $\bar{F}_{xy}^{(7)}$ is proportional to the skyrmion density of the vector field $\bm{n}(\bm{r})$ defined as $\rho_{\mathrm{AFM-SkX}}=(1/4\pi V)\int \mathrm{d}^{2}\bm{r}\ \bm{n}(\bm{r})\cdot(\partial_{x}\bm{n}(\bm{r})\times\partial_{y}\bm{n}(\bm{r}))$, and the other two terms are from the coupling between the DM interaction and vector field $\bm{n}(\bm{r})$. Returning to the original lattice, finite $F_{xy}(\bm{r})$ generates a uniformly distributed flux $\Phi=\bar{F}_{xy}^{(7)}\lambda_{7}v'/2$ as shown in Fig.~\ref{fig:skx}(d), which breaks the effective time-reversal symmetry and contributes to the magnon thermal Hall effect.

We finally comment on SU(3) gauge redundancy. The SU(3) gauge transformation is given by the local rotation of the internal degrees of freedom, namely $\hat{\bm{\gamma}}(\bm{r})\to W(\bm{r})\hat{\bm{\gamma}}(\bm{r})$ and $T(\bm{r})\to W(\bm{r})T(\bm{r})W^{\dagger}(\bm{r})-i(\bm{\nabla}W(\bm{r}))W^{\dagger}(\bm{r})$ with $W\in\mathrm{SU(3)}$. The latter transformation corresponds to the rotation of the vector field $F_{xy}^{(\alpha)}(\bm{r})$ in the eight-dimensional space, indicating that the length of the vector field $\sqrt{\sum_{\alpha=1}^{8}(F_{xy}^{(\alpha)}(\bm{r}))^{2}}$ is gauge invariant. Therefore, the emergence of the finite $F_{xy}(\bm{r})$ is not an artifact of the gauge choice and characterizes the effect of the spin texture in the AFM-SkXs.

\emph{\textbf{Thermal Hall conductivity.---}}To demonstrate that the finite field strength from the AFM-SkX spin texture is responsible for the magnon thermal Hall effect, we calculate the thermal Hall conductivity by the linear spin-wave theory. We first determine the spin configurations in the classical ground state by the mean-field approximation. Applying the Holstein-Primakoff transformation, the spin model~(\ref{eq:H}) is reduced to the quadratic magnon Hamiltonian $\hat{\mathcal{H}}_{\mathrm{mag}}=(1/2)\sum_{\bm{k}}\hat{\Phi}_{\bm{k}}^{\dagger}H_{\mathrm{BdG}}(\bm{k})\hat{\Phi}_{\bm{k}}$ up to the constant term, where $\hat{\Phi}_{\bm{k}}=(\hat{b}_{\bm{k},1},\cdots,\hat{b}_{\bm{k},N_{\mathrm{m}}},\hat{b}_{-\bm{k},1}^{\dagger},\cdots,\hat{b}_{-\bm{k},N_{\mathrm{m}}}^{\dagger})^{T}$, $\hat{b}_{\bm{k},s}$ ($\hat{b}_{\bm{k},s}$) is the magnon annihilation (creation) operator with wave vector $\bm{k}$ and magnetic sublattice index $s$, $N_{\mathrm{m}}$ is the number of sites in a magnetic unit cell, and $H_{\mathrm{BdG}}(\bm{k})$ is the $2N_{\mathrm{m}}\times2N_{\mathrm{m}}$ Hermitian matrix containing systems' details. The Bogoliubov transformation, which preserves the bosonic commutation relations, introduces new bosonic operators $\hat{\Phi}_{\bm{k}}=P(\bm{k})\hat{\Psi}_{\bm{k}}$, where $\hat{\Psi}_{\bm{k}}=(\hat{\beta}_{\bm{k},1},\cdots,\hat{\beta}_{\bm{k},N_{\mathrm{m}}},\hat{\beta}_{-\bm{k},1}^{\dagger},\cdots,\hat{\beta}_{-\bm{k},N_{\mathrm{m}}}^{\dagger})^{T}$ and $P(\bm{k})$ is the paraunitary matrix that diagonalizes $H_{\mathrm{BdG}}(\bm{k})$~\cite{colpa1978pa}. The magnon Hamiltonian then takes the diagonal form $\hat{\mathcal{H}}_{\mathrm{mag.}}=\sum_{\bm{k}}\sum_{n=1}^{N_{\mathrm{m}}}\varepsilon_{n}(\bm{k})(\hat{\beta}_{\bm{k},n}^{\dagger}\hat{\beta}_{\bm{k},n}+1/2)$ with the $n$th magnon band $\varepsilon_{n}(\bm{k})$. See Supplemental Material for more details~\cite{supp}.

\begin{figure}[t]
    \includegraphics[width=85mm]{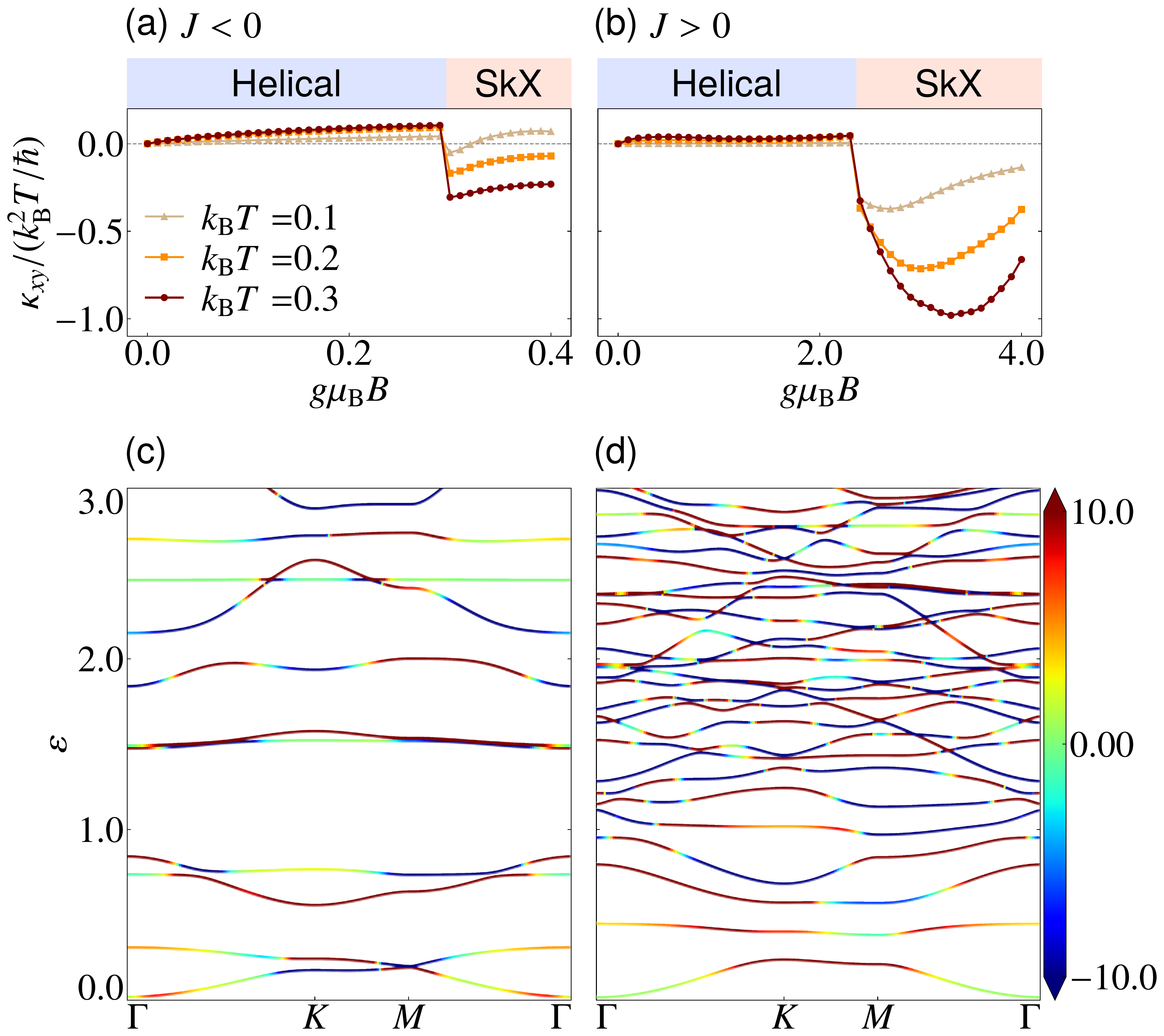}
    \caption{\textbf{Magnon bands and thermal Hall conductivity.} (a), (b) Magnetic-field dependence of the thermal Hall conductivity at $k_{\mathrm{B}}T=0.1, 0.2, 0.3$ for (a) $J<0$ and (b) $J>0$. (c), (d) Magnon bands along the symmetric line in the Brillouin zone for the (c) FM-SkX phase with $g\mu_{\mathrm{B}}B=0.3$ and (d) AFM-SkX phase with $g\mu_{\mathrm{B}}B=3.0$. The color density represents the Berry curvature. We set $|J|S=1.0$ and $D/|J|=1.0$ ($0.5$) for $J<0$ ($J>0$).}
    \label{fig:hall}
\end{figure}

The thermal Hall conductivity in the linear response regime is given by~\cite{matsumoto2011prb,matsumoto2011prl,matsumoto2014prb}
\begin{align}
    \kappa_{xy}=-\frac{k_{\mathrm{B}}^{2}T}{\hbar}\int_{\mathrm{BZ}}\frac{\mathrm{d}^{2}\bm{k}}{(2\pi)^{2}}\sum_{n=1}^{N_{\mathrm{m}}}c_{2}[f(\varepsilon_{n}(\bm{k}))]\Omega_{xy}^{(n)}(\bm{k}),
\end{align}
where $c_{2}[x]=\int_{0}^{x}dt[\ln(1+t^{-1})]^{2}$, $f(\varepsilon)=1/[\exp(\varepsilon/k_{\mathrm{B}}T)-1]$ is the Bose distribution function, and $\Omega_{xy}^{(n)}(\bm{k})=-2\mathrm{Im}[\{\partial_{k_{x}}T^{\dagger}(\bm{k})\}\Sigma^{z}\{\partial_{k_{y}}T(\bm{k})\}]_{n,n}$ is the Berry curvature of $n$th magnon bands, and $\Sigma^{z}=\tau^{z}\otimes I_{N_{\mathrm{m}}\times N_{\mathrm{m}}}$ with $N_{\mathrm{m}}\times N_{\mathrm{m}}$ unit matrix $I_{N_{\mathrm{m}}\times N_{\mathrm{m}}}$. Figures~\ref{fig:hall}(a) and~\ref{fig:hall}(b) show the magnetic-field dependence of the thermal Hall conductivity with (a) $JS=-1.0$ and $D/|J|=1.0$ and (b) $JS=1.0$ and $D/|J|=0.5$~\footnote{We choose different values of $D/|J|$ for the FM-SkX and AFM-SkX so that the spin configurations fit within a numerically tractable magnetic unit cell. See Supplemental Material for details~\cite{supp}.}. In both the FM and AFM cases, $\kappa_{xy}$ remains very small in the helical phases and becomes very large upon entering the skyrmion phases. This indicates that FM-SkX (AFM-SkX) spin texture is the primary origin of the magnon thermal Hall effect, supporting our field-theoretical results.

Finally, we discuss how the distinct gauge structures of the FM-SkXs and AFM-SkXs possibly influence the thermal Hall conductivity. While field-theoretical approaches often lack quantitative precision, they may still capture the overall tendency in the thermal Hall conductivity. In the FM-SkXs, the finite skyrmion density produces the uniform component of the effective magnetic field and leads to magnon Landau Levels, whose spacing is proportional to the field~\cite{hoogdalem2013prb,weber2022science}. A large skyrmion density then pushes most Landau levels out of the low-energy region. Since magnons obey Bose statistics, the thermal Hall conductivity decreases with increasing the skyrmion density, in contrast to electronic systems~\cite{kim2019prl}. In the AFM-SkXs, the amplitude of the field strength from the spin texture is also proportional to the skyrmion density of the vector field $\bm{n}(\bm{r})$, whose amplitude is identical to the skyrmion density of the FM-SkX on each sublattice. However, the prefactor $1/8$ and additional terms from the DM interaction in Eq.~(\ref{eq:F7}) imply that the effective magnetic field in the AFM-SkXs tends to be smaller than that in the FM-SkXs. Therefore, we roughly expect that the thermal Hall conductivity of the AFM-SkXs will be larger than that of the FM-SkXs.

Figures~\ref{fig:hall}(c) and~\ref{fig:hall}(d) show the magnon bands and Berry curvatures in the FM-SkX phase with $JS=-1.0$, $D/|J|=1.0$, and $g\mu_{\mathrm{B}}B=0.3$ and AFM-SkX phase with $JS=1.0$, $D/|J|=0.5$, and $g\mu_{\mathrm{B}}B=3.0$ along the symmetric line in the Brillouin zone. Here, $\Gamma$, $K$, and $M$ points in Figs.~\ref{fig:hall}(c) and~\ref{fig:hall}(d) correspond to $\bm{k}=(0.0)$, $(4\pi/3L_{\mathrm{m}},0)$, and $(2\pi/3L_{\mathrm{m}},\pi/\sqrt{3}L_{\mathrm{m}})$, where $L_{\mathrm{m}}=9$ ($7$) for the FM-SkX (AFM-SkX) and $N_{\mathrm{m}}=L_{\mathrm{m}}^{2}$. The amplitude of the skyrmion density is numerically estimated as $|\rho_{\mathrm{FM-SkX}}|\simeq8.9\times10^{-3}$ for the FM-SkX phase with $g\mu_{\mathrm{B}}B=0.3$ and $\rho_{\mathrm{AFM-SkX}}\simeq1.2\times10^{-2}$ for the AFM-SkX phase with $g\mu_{\mathrm{B}}=3.0$~\cite{supp}. Despite $|\rho_{\mathrm{AFM-SkX}}|>|\rho_{\mathrm{FM-SkX}}|$, the magnon bands in the AFM-SkX are narrower and more densely spaced than those in the FM-SkX, suggesting a larger thermal Hall conductivity in the AFM-SkX. The thermal Hall conductivity in the AFM-SkX phase (Fig.~(\ref{fig:hall}(b))) is indeed larger than that in the FM-SkX (Fig.~\ref{fig:hall}(a)), supporting our conjecture. A more quantitative understanding of the relationship between the field strength and thermal Hall conductivity in AFM-SkXs phase remains for future work.

\emph{\textbf{Conclusions and outlook.---}}We have shown that low-energy excitations in the AFM-SkXs on the triangular lattice are effectively governed by the magnons with the SU(3) gauge field. We have also demonstrated that the finite commutation relation of the SU(3) gauge field breaks the effective time-reversal symmetry and underpins a measurable magnon thermal Hall effect. The emergence of the thermal Hall conductivity in the AFM-SkX phase has also been demonstrated by the linear spin-wave theory. The relation between the gauge structures and magnitude of the thermal Hall conductivity has also been discussed. In contrast to earlier field-theoretical approaches that rely on the Lagrangian formalism and involve integrating out uniform magnetization components~\cite{dombre1989prb,pradenas2024prb}, our formulation is simpler and proceeds in a straightforward manner, analogous to the standard Holstein-Primakoff transformation. We have explicitly derived the connection between the AFM-SkX spin texture and the SU(3) gauge field, and we have further clarified how its non-commutative nature leads to the effective time-reversal symmetry breaking.

There are several promising directions for future work. For example, it would be interesting to explore fictitious gauge fields in different classes of skyrmions~\cite{borge2021phyrep}, including two-sublattice AFM-SkXs~\cite{zhang2016scirep,gobel2017prb,hayami2023jpsj}. Our framework is readily applicable to these systems. The detailed analysis of the relation between gauge structures and thermal Hall conductivity is left for future research. Elucidating how an underlying gauge structure qualitatively changes other physical quantities would also offer an interesting research direction.

\emph{\textbf{Acknowledgments.---}}M.K. thanks C. Hotta, Karlo Penc, and Levente R\'{o}zsa for discussion. M.K. was supported by JSPS Overseas Research Fellowship.

\emph{\textbf{Data availability.---}} Data and codes are avaialbe on Zenodo upon reasonable request~\cite{zenodo}.

\bibliography{biblio}

\newpage
\leavevmode \newpage
\onecolumngrid
\appendix
\begin{center}
\textbf{Supplemental Material\\ \papertitle}\\
\vspace{10pt}
Masataka Kawano\\
\vspace{5pt}
\textit{\small{\tum}}\\
\vspace{10pt}
\end{center}
\twocolumngrid

\section{Derivation of the equation of motion}
The equation of motion for $\hat{\Psi}(\bm{r},t)=(\gamma_{1}(\bm{r}),\gamma_{1}^{\dagger}(\bm{r}),\gamma_{2}(\bm{r}),\gamma_{2}^{\dagger}(\bm{r}),\gamma_{3}(\bm{r}),\gamma_{3}^{\dagger}(\bm{r}))^{T}$ is calculated as
\begin{align}
    i\hbar\partial_{t}\hat{\Psi}(\bm{r},t)=(I\otimes\tau^{z})H(\bm{r})\hat{\Psi}(\bm{r},t).
\end{align}
By applying this equation twice, we have
\begin{align}
    \hbar^{2}\partial_{t}^{2}\hat{\Psi}(\bm{r},t)=-(I\otimes\tau^{z})H(\bm{r})(I\otimes\tau^{z})H(\bm{r})\hat{\Psi}(\bm{r},t).
    \label{eq:eom-twice}
\end{align}
To evaluate the $6\times6$ matrix on the right-hand side of Eq.~(\ref{eq:eom-twice}), we first divide $H(\bm{r})$ defined in Eq.~(\ref{eq:Hr}) into two parts
\begin{align}
    H(\bm{r})=H_{0}+H_{\nabla^{2}}(\bm{r})+H_{G}(\bm{r})+H_{B},
\end{align}
where
\begin{align}
    H_{0}&=\frac{9\sqrt{2}}{4}JSI\otimes(\tau^{0}+\tau^{x}),\\
    H_{\nabla^{2}}(\bm{r})&=\frac{3\sqrt{2}}{8}JSa^{2}\tilde{I}\otimes\tau^{x}\nabla^{2},\\
    H_{G}(\bm{r})&=-i\frac{3\sqrt{2}}{8}JSa^{2}G(\bm{r})\cdot\bm{\nabla},\\
    H_{B}&=g\mu_{\mathrm{B}}BM.
\end{align}
The $6\times6$ matrix in the right-hand side of Eq.~(\ref{eq:eom-twice}) is approximated as
\begin{align}
    &(I\otimes\tau^{z})H(\bm{r})(I\otimes\tau^{z})H(\bm{r})\nonumber\\
    &\simeq(I\otimes\tau^{z})H_{0}(I\otimes\tau^{z})H_{0}\nonumber\\
    &\quad+(I\otimes\tau^{z})H_{0}(I\otimes\tau^{z})[H_{\nabla^{2}}(\bm{r})+H_{G}(\bm{r})+H_{B}]\nonumber\\
    &\quad+(I\otimes\tau^{z})[H_{\nabla^{2}}(\bm{r})+H_{G}(\bm{r})+H_{B}](I\otimes\tau^{z})H_{0}.
    \label{eq:tauzHtauzH}
\end{align}
Here, we ignore higher-order derivative terms, which only give negligible contributions at low energies. We also assume that $g\mu_{\mathrm{B}}BM$ is small. The first term in Eq.~(\ref{eq:tauzHtauzH}) vanishes
\begin{align}
    (I\otimes\tau^{z})H_{0}(I\otimes\tau^{z})H_{0}&=\frac{81}{8}(JS)^{2}I\otimes[(\tau^{0}-\tau^{x})(\tau^{0}+\tau^{x})]\nonumber\\
    &=0,
\end{align}
where $\tau^{z}\tau^{x}\tau^{z}=-\tau^{x}$. One can calculate the other terms in Eq.~(\ref{eq:tauzHtauzH}) in a similar manner. For example, the terms including $H_{\nabla^{2}}(\bm{r})$ are calculated as
\begin{align}
    &(I\otimes\tau^{z})H_{0}(I\otimes\tau^{z})H_{\nabla^{2}}(\bm{r})+(I\otimes\tau^{z})H_{\nabla^{2}}(\bm{r})(I\otimes\tau^{z})H_{0}\nonumber\\
    &=\frac{27}{16}(JSa)^{2}\nabla^{2}\tilde{I}\otimes[(\tau^{0}-\tau^{x})\tau^{x}-\tau^{x}(\tau^{0}+\tau^{x})]\nonumber\\
    &=-\frac{27}{8}(JSa)^{2}\tilde{I}\otimes\tau^{0}.
\end{align}
We remark that the term proportional to $\bm{\xi}(\bm{r})$ does not contribute since
\begin{align}
    &(I\otimes\tau^{z})H_{0}(I\otimes\tau^{z})\left[\bm{\xi}(\bm{r})\cdot\bm{\nabla}\lambda_{7}\otimes(\tau^{0}+\tau^{x})\right]\nonumber\\
    &\qquad+(I\otimes\tau^{z})\left[\bm{\xi}(\bm{r})\cdot\bm{\nabla}\lambda_{7}\otimes(\tau^{0}+\tau^{x})\right](I\otimes\tau^{z})H_{0}\nonumber\\
    &\propto\bm{\xi}(\bm{r})\cdot\bm{\nabla}\lambda_{7}(\tau^{0}-\tau^{x})(\tau^{0}+\tau^{x})\nonumber\\
    &=0.
\end{align}
Most of the terms in $(I\otimes\tau^{z})H(\bm{r})(I\otimes\tau^{z})H(\bm{r})$ become diagonal. Some off-diagonal terms also appear, but they do not contribute to the equation of motion at low energies. For example, $\sqrt{2}\bm{A}_{\mathrm{s}}(\bm{r})\lambda_{6}\otimes\tau^{z}$ in $\bm{G}(\bm{r})$ gives
\begin{align}
    &(I\otimes\tau^{z})H_{0}(I\otimes\tau^{z})\left[\sqrt{2}\bm{A}_{\mathrm{s}}(\bm{r})\lambda_{6}\otimes\tau^{z}\right]\nonumber\\
    &\qquad+(I\otimes\tau^{z})\left[\sqrt{2}\bm{A}_{\mathrm{s}}(\bm{r})\lambda_{6}\otimes\tau^{z}\right](I\otimes\tau^{z})H_{0}\nonumber\\
    &\propto\bm{A}_{\mathrm{s}}(\bm{r})\lambda_{6}\otimes[(\tau^{0}-\tau^{x})\tau^{z}+\tau^{z}(\tau^{0}+\tau^{x})]\nonumber\\
    &\propto\bm{A}_{\mathrm{s}}(\bm{r})\lambda_{6}\otimes(\tau^{z}+i\tau^{y})\nonumber\\
    &=\bm{A}_{\mathrm{s}}(\bm{r})\lambda_{6}\otimes\tau^{z}+O((\bm{A}_{\mathrm{s}}(\bm{r}))^{2}),
\end{align}
and the off-diagonal term contributes at second order in $\bm{A}_{\mathrm{s}}(\bm{r})$ as a result of diagonalization with a paraunitary matrix. Summing all terms in Eq.~(\ref{eq:tauzHtauzH}) leads to the equation of motion~(\ref{eq:eom_gamma}).

\section{Classical ground state}
To perform the linear spin-wave theory, we need to find the spin configuration in the classical ground state. The classical ground state of the spin model (\ref{eq:H}) in the main text is obtained in Refs.~\cite{diaz2019prl,diaz2020prr,mook2020prr} by the simulated annealing and subsequent time evolution. Here, we employ a standard mean-field approximation, which is enough to reproduce the results in Refs.~\cite{diaz2019prl,diaz2020prr,mook2020prr}.

We first employ the Luttinger-Tisza method to the spin model (\ref{eq:H}) with $B=0$ to construct an initial spin configuration for the mean-field approximation. We consider the energy minimization within a product state $\ket{\Phi_{\mathrm{MF}}}=\bigotimes_{i=1}^{N}\ket{\bm{m}_{i}}$, where $\ket{\bm{m}_{i}}$ is the spin coherent state satisfying $\braket{\bm{m}_{i}|\hat{\bm{S}}_{i}|\bm{m}_{i}}=S\bm{m}_{i}$ with three-dimensional unit vector $\bm{m}_{i}$ representing the direction of spin at site $i$ and $N$ is the number of sites. The expectation value of the energy, $E_{\mathrm{MF}}^{(B=0)}=\braket{\Phi_{\mathrm{MF}}|\hat{\mathcal{H}}|\Phi_{\mathrm{MF}}}$, is calculated as
\begin{align}
    E_{\mathrm{MF}}^{(B=0)}&=S^{2}\sum_{\braket{i,j}}\left[J\bm{m}_{i}\cdot\bm{m}_{j}+\bm{D}_{i,j}\cdot(\bm{m}_{i}\times\bm{m}_{j})\right].
    \label{eq:E_MF_B=0}
\end{align}
In the Luttinger-Tisza method, we relax the local constraint $|\bm{m}_{i}|^{2}=1$ to a global one $\sum_{i=1}^{N}|\bm{m}_{i}|^{2}=N$ and minimize the following function instead of $E_{\mathrm{MF}}^{(B=0)}$
\begin{align}
    E_{\mathrm{LT}}=E_{\mathrm{MF}}^{(B=0)}-\lambda S^{2}\left(\sum_{i=1}^{N}|\bm{m}_{i}|^{2}-N\right),
    \label{eq:E_LT}
\end{align}
where $\lambda$ is the Lagrange multiplier. We introduce the Fourier transformation of $\bm{m}_{\bm{r}}$ as
\begin{align}
    \bm{m}_{\bm{r}}=\sum_{\bm{q}}\bm{m}(\bm{q})\mathrm{e}^{i\bm{q}\cdot\bm{r}},
\end{align}
which transforms Eq.~(\ref{eq:E_LT}) to
\begin{align}
    E_{\mathrm{LT}}=\lambda NS^{2}+NS^{2}\sum_{\bm{q}}\bm{m}^{\dagger}(\bm{q})[F(\bm{q})-I]\bm{m}(\bm{q}).
\end{align}
Here, $F(\bm{q})$ is the $3\times3$ Hermitian matrix
\begin{align}
    F(\bm{q})=
    \begin{pmatrix}
        J\sum_{\mu}\cos q_{\mu} & 0 & iDf(\bm{q})\\
        0 & J\sum_{\mu}\cos q_{\mu} & iDg(\bm{q})\\
        -iDf(\bm{q}) & -iDg(\bm{q}) & J\sum_{\mu}\cos q_{\mu}
    \end{pmatrix}
    ,
\end{align}
where $q_{\mu}=\bm{q}\cdot\bm{e}_{\mu}$, $\bm{e}_{1}=(1,0)$, $\bm{e}_{2}=(-1/2,\sqrt{3}/2)$, $\bm{e}_{3}=(-1/2,-\sqrt{3}/2)$, and
\begin{align}
    f(\bm{q})&=-\sin q_{1}+\frac{1}{2}(\sin q_{2}+\sin q_{3}),\\
    g(\bm{q})&=\frac{\sqrt{3}}{2}(-\sin q_{2}+\sin q_{3}).
\end{align}
The eigenvalue of $F(\bm{q})$ is calculated as
\begin{align}
    w_{1}(\bm{q})&=J\sum_{\mu}\cos q_{\mu}-|D|\sqrt{f^{2}(\bm{q})+g^{2}(\bm{q})},\\
    w_{2}(\bm{q})&=J\sum_{\mu}\cos q_{\mu},\\
    w_{3}(\bm{q})&=J\sum_{\mu}\cos q_{\mu}+|D|\sqrt{f^{2}(\bm{q})+g^{2}(\bm{q})},
\end{align}
The wave vector $\bm{Q}$ that minimizes the minimum eigenvalue $w_{1}(\bm{q})$ characterizes the spin configuration in the classical ground state. The minimum eigenvalue $w_{1}(\bm{q})$ takes the minimum value at $\bm{Q}_{\mu}=Q\bm{e}_{\mu}$, where $Q$ ($0\leq Q\leq4\pi/3$) satisfies the following relation,
\begin{align}
    J\left(\sin Q+\sin\frac{Q}{2}\right)+|D|\left(\cos Q+\frac{1}{2}\cos\frac{Q}{2}\right)=0.
    \label{eq:find_Q}
\end{align}
Since $\bm{m}_{\bm{r}}\in\mathbb{R}^{3}$, $w_{1}(\bm{q})$ also takes the minimum value at $-\bm{Q}_{\mu}$. To find the spin configuration, we also need to find the eigenvector for $w_{1}(\bm{Q}_{\mu})$, which is calculated as
\begin{align}
    \bm{m}_{1}(\bm{Q}_{\mu}) = \frac{\mathrm{e}^{i\varphi(\bm{Q}_{\mu})}}{\sqrt{2}}
    \begin{pmatrix}
        -if(\bm{Q}_{\mu})/\sqrt{f^{2}(\bm{Q}_{\mu})+g^{2}(\bm{Q}_{\mu})}\\
        -ig(\bm{Q}_{\mu})/\sqrt{f^{2}(\bm{Q}_{\mu})+g^{2}(\bm{Q}_{\mu})}\\
        1
    \end{pmatrix}
    ,
\end{align}
where $\varphi(\bm{Q}_{\mu})$ is the arbitrary phase factor, which satisfies $\varphi(\bm{Q}_{\mu})=-\varphi(-\bm{Q}_{\mu})=\varphi_{\mu}$ since $\bm{m}_{\bm{r}}\in\mathbb{R}$. The solution with single $\bm{Q}_{\mu}$ is then given by
\begin{align}
    \bm{m}_{\bm{r}} &= \frac{1}{\sqrt{2}}\left(\bm{m}_{1}(\bm{Q}_{\mu})\mathrm{e}^{i\bm{Q}_{\mu}\cdot\bm{r}}+\bm{m}_{1}(-\bm{Q}_{\mu})\mathrm{e}^{-i\bm{Q}_{\mu}\cdot\bm{r}}\right),
\end{align}
and the explicit calculation of $f(\bm{Q}_{\mu})$ and $g(\bm{Q}_{\mu})$ leads to three helical ground states,
\begin{align}
   \bm{m}_{\bm{r}} = -\bm{e}^{\mu}\sin(\bm{Q}_{\mu}\cdot\bm{r}+\varphi_{\mu})+\bm{e}^{Z}\cos(\bm{Q}_{\mu}\cdot\bm{r}+\varphi_{\mu}),
   \label{eq:mr_helical}
\end{align}
where $\bm{e}^{1}=(1,0,0)$, $\bm{e}^{2}=(-1/2,\sqrt{3}/2,0)$, $\bm{e}^{3}=(-1/2,-\sqrt{3}/2,0)$, and $\bm{e}^{Z}=(0,0,1)$. This solution satisfies the local constraint, $|\bm{m}_{\bm{r}}|^{2}=1$. As a initial state, we consider the helical state~(\ref{eq:mr_helical}) and the linear combination of three helical states with the uniform component
\begin{align}
    \bm{m}_{\bm{r}}=A\sum_{\mu=1}^{3}\left(\bm{e}^{\mu}\sin(\bm{Q}_{\mu}\cdot\bm{r})-\bm{e}^{Z}\cos(\bm{Q}_{\mu}\cdot\bm{r})\right)+M\bm{e}^{Z},
    \label{eq:mr_init}
\end{align}
where $A>0$ and $M>0$ are determined by the normalization condition $|\bm{m}_{i}|^{2}=1$. We also assume that the spins at the center of the skyrmions point in the opposite direction of the magnetic field, which leads to $\varphi_{\mu}=\pi$ ($\mu=1.2.3$). We remark that for $J<0$, the helical states with a different wave vector
\begin{align}
   \bm{m}_{\bm{r}} = \tilde{\bm{e}}^{\mu}\sin(\tilde{\bm{Q}}_{\mu}\cdot\bm{r}+\tilde{\varphi}_{\mu})+\bm{e}^{Z}\cos(\tilde{\bm{Q}}_{\mu}\cdot\bm{r}+\tilde{\varphi}_{\mu}),
   \label{eq:mr_helical_fm}
\end{align}
or the skyrmion states built from these linear combinations
\begin{align}
    \bm{m}_{\bm{r}}=A\sum_{\mu=1}^{3}\left(\tilde{\bm{e}}^{\mu}\sin(\tilde{\bm{Q}}_{\mu}\cdot\bm{r})-\bm{e}^{Z}\cos(\tilde{\bm{Q}}_{\mu}\cdot\bm{r})\right)+M\bm{e}^{Z},
    \label{eq:mr_init_fm}
\end{align}
yields almost the same classical ground-state energy, where $\tilde{\bm{Q}}_{\mu}=Q\tilde{\bm{e}}_{\mu}$, $\tilde{\bm{e}}_{1}=(\sqrt{3}/2,1/2)$, $\tilde{\bm{e}}_{2}=(-\sqrt{3}/2,1/2)$, $\tilde{\bm{e}}_{3}=(0,-1)$, and $\tilde{\varphi}_{\mu}=\pi$ ($\mu=1,2,3$). Since the system with $J<0$ prefers the states with $\tilde{\bm{Q}}_{\mu}$ in Refs.~\cite{diaz2020prr,mook2020prr}, we employ Eqs.~(\ref{eq:mr_helical_fm}) and~(\ref{eq:mr_init_fm}) as the initial states.

\begin{figure*}[t]
    \centering
    \includegraphics[width=175mm]{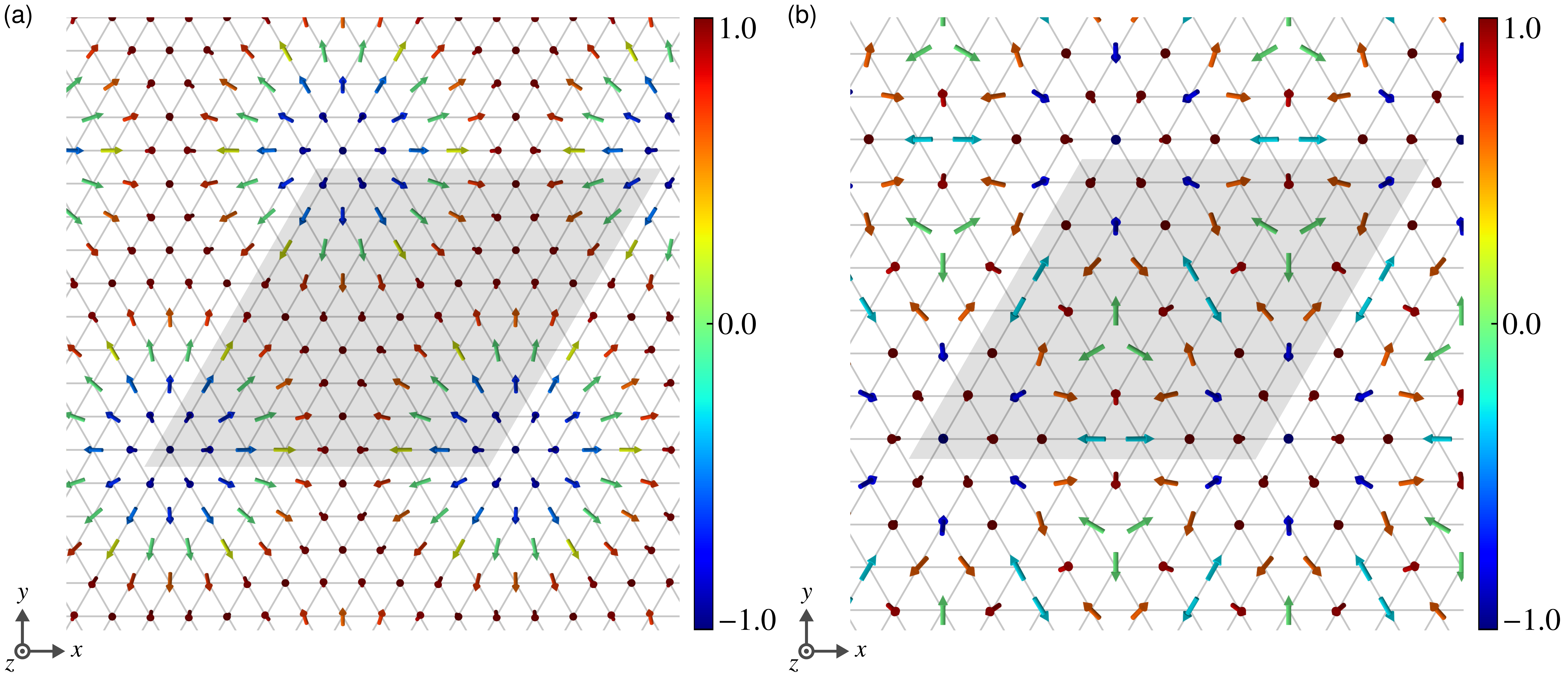}
    \caption{\textbf{Spin configuration in the classical ground state.} (a), (b) Spin texture of the classical ground state in the (a) FM-SkX phase and (b) AFM-SkX phase. The magnetic unit cell is shaded in light gray. The color represents the $z$-component of the spins.}
    \label{fig:cgs}
\end{figure*}

The wave number $Q$ depends on $D/|J|$ and is incommensurate in general. Here, we carefully choose $D/|J|$ such that the magnetic ordering is compatible with the size of the unit cell that is tractable in numerical calculations. For a given shape of the unit cell $N_{\mathrm{m}}=L_{\mathrm{m}}\times L_{\mathrm{m}}$ (see Fig.~\ref{fig:cgs}(a)), possible wave number $Q$ is given by $Q=4\pi n_{Q}/L_{\mathrm{m}}$ ($n_{Q}=1,2,\ldots,[L_{\mathrm{m}}/3]$). We chose $D/|J|$ and $L_{\mathrm{m}}$ such that $n_{Q}$ approximately takes integer value. For the FM-SkX (AFM-SkX), we choose $D/|J|=1.0$ ($0.5$), which lead to $L_{\mathrm{m}}=9$ ($7$) and $n_{Q}=1$ ($2$) from Eq.~(\ref{eq:find_Q}).

The mean-field approximation is formulated based on a variational principle; we minimize the expectation value of the energy within a product state
\begin{align}
    E_{\mathrm{MF}}&=E_{\mathrm{MF}}^{(B=0)}-g\mu_{\mathrm{B}}BS\sum_{i}m_{i}^{z}.
    \label{eq:E_MF}
\end{align}
We introduce the mean-field Hamiltonian, $\hat{\mathcal{H}}_{\mathrm{MF}}$, which shares same expectation value of the energy, $E_{\mathrm{MF}}=\braket{\Phi_{\mathrm{MF}}|\hat{\mathcal{H}}_{\mathrm{MF}}|\Phi_{\mathrm{MF}}}$, as
\begin{align}
    \hat{\mathcal{H}}_{\mathrm{MF}}=E_{\mathrm{MF}}^{(B=0)}-\sum_{i=1}^{N}\bm{h}_{i}^{(\mathrm{MF})}\cdot\hat{\bm{S}}_{i}.
\end{align}
where $\bm{h}_{i}^{(\mathrm{MF})}$ is the effective magnetic field including the effect of surrounding spins
\begin{align}
    \bm{h}_{i}^{(\mathrm{MF})}=S\sum_{j\in\Lambda_{\mathrm{n.n.}i}}\left[-J\bm{m}_{j}+\bm{D}_{i,j}\times\bm{m}_{j}\right]+g\mu_{\mathrm{B}}B\bm{e}^{Z},
\end{align}
and $\Lambda_{\mathrm{n.n.}i}$ is the set of sites nearest neighbors to site $i$. The spin configuration in the mean-field ground state is determined by the self-consistent equation
\begin{align}
    \bm{m}_{i}=\frac{\bm{h}_{i}^{(\mathrm{MF})}}{|\bm{h}_{i}^{(\mathrm{MF})}|}.
    \label{eq:self-consistent-eq}
\end{align}
Figure~\ref{fig:cgs}(a) and~\ref{fig:cgs}(b) show the spin configuration in the classical ground state obtained from the self-consistent equation~(\ref{eq:self-consistent-eq}), which reproduces the spin configurations in Ref.~\cite{diaz2019prl,diaz2020prr}.

The FM-SkX phase is characterized by a skyrmion number
\begin{align}
    N_{\mathrm{FM-SkX}}=\frac{1}{4\pi}\sum_{\braket{i,j,k}}\Omega_{i,j,k},
\end{align}
where $\braket{i,j,k}$ denotes the set of sites on an elementary triangle with counterclockwise indices and
\begin{align}
    \Omega_{i,j,k}=2\tan^{-1}\frac{\bm{m}_{i}\cdot(\bm{m}_{j}\times\bm{m}_{k})}{(1+\bm{m}_{i}\cdot\bm{m}_{j}+\bm{m}_{j}\cdot\bm{m}_{k}+\bm{m}_{k}\cdot\bm{m}_{i})},
\end{align}
is the solid angle subtended by three unit vectors $\bm{m}_{i}$, $\bm{m}_{j}$, and $\bm{m}_{k}$~\cite{berg1981npb}. In the continuum limit $\bm{m}_{i}\to v\bm{m}(\bm{r})$, $N_{\mathrm{FM-SkX}}$ is reduced to
\begin{align}
    N_{\mathrm{FM-SkX}}\to\frac{1}{4\pi}\int\mathrm{d}^{2}\bm{r}\ \bm{m}(\bm{r})\cdot(\partial_{x}\bm{m}(\bm{r})\times\partial_{y}\bm{m}(\bm{r}))
\end{align}
The skyrmion density $\rho_{\mathrm{FM-SkX}}$ in the main text corresponds to $N_{\mathrm{FM-SkX}}/N$. The AFM phase is characterized by the skyrmion number defined on each sublattice, $N_{\mathrm{FM-SkX}}^{(\ell)}$ ($\ell=1,2,3$). We define the skyrmion number for the AFM-SkX as $N_{\mathrm{AFM-SkX}}=(1/3)\sum_{\ell}N_{\mathrm{FM-SkX}}^{(\ell)}$. The amplitude of the skyrmion density of the vector field $\bm{n}(\bm{r})$, namely $\rho_{\mathrm{AFM-SkX}}$ in the main text, corresponds to $|N_{\mathrm{AFM-SkX}}/N|$.

\section{Linear spin-wave theory}
We apply the linear spin-wave theory to find the magnon bands and eigenstates. The Holstein-Primakoff transformation within the linear spin-wave approximation is given by
\begin{align}
    \hat{\bm{S}}_{i} \simeq \sqrt{S}\left(\hat{b}_{i}\bm{e}_{i}^{-}+\hat{b}_{i}^{\dagger}\bm{e}_{i}^{+}\right) + \left(S-\hat{b}_{i}^{\dagger}\hat{b}_{i}\right)\bm{m}_{i},
\end{align}
where $\bm{e}_{i}^{\pm}=(\bm{e}_{i}^{x}\pm\bm{e}_{i}^{y})/\sqrt{2}$ and $\bm{e}_{i}^{x}$, $\bm{e}_{i}^{y}$, $\bm{m}_{i}$ form the local orthogonal basis. Within the linear spin-wave approximation, the spin Hamiltonian (\ref{eq:H}) is transformed as
\begin{align}
    \hat{\mathcal{H}}&\simeq E_{\mathrm{MF}}+\sum_{\alpha}\sum_{\bm{r}\in\Lambda_{\alpha}}\varepsilon_{\alpha}\hat{b}_{\bm{r}}^{\dagger}\hat{b}_{\bm{r}}\nonumber\\
    &\quad+\sum_{\alpha}\sum_{\bm{r}\in\Lambda_{\alpha}}\sum_{\mu}\left(t_{\alpha,\alpha_{\mu}}\hat{b}_{\bm{r}}^{\dagger}\hat{b}_{\bm{r}+\bm{e}_{\mu}}+\Delta_{\alpha,\alpha_{\mu}}\hat{b}_{\bm{r}}^{\dagger}\hat{b}_{\bm{r}+\bm{e}_{\mu}}^{\dagger}+\mathrm{H.c.}\right),
    \label{eq:Hmag_real}
\end{align}
where $\Lambda_{\alpha}$ is the set of sites with sublattice index $\alpha$, $\alpha_{\mu}$ is the sublattice index for $\bm{r}+\bm{e}_{\mu}$ with $\bm{r}\in \Lambda_{\alpha}$, and
\begin{align}
    \varepsilon_{\alpha}&=-2S\sum_{\mu}\left(J\bm{m}_{\alpha}\cdot\bm{m}_{\alpha_{\mu}}+\bm{D}_{\mu}\cdot(\bm{m}_{\alpha}\times\bm{m}_{\alpha_{\mu}})\right)\nonumber\\
    &\quad+g\mu_{\mathrm{B}}Bm_{\alpha}^{z}+\Delta h,\\
    t_{\alpha,\alpha_{\mu}}&=JS\bm{e}_{\alpha}^{+}\cdot\bm{e}_{\alpha_{\mu}}^{-}+S\bm{D}_{\mu}\cdot(\bm{e}_{\alpha}^{+}\times\bm{e}_{\alpha_{\mu}}^{-}),\\
    \Delta_{\alpha,\alpha_{\mu}}&=JS\bm{e}_{\alpha}^{+}\cdot\bm{e}_{\alpha_{\mu}}^{+}+S\bm{D}_{\mu}\cdot(\bm{e}_{\alpha}^{+}\times\bm{e}_{\alpha_{\mu}}^{+}),
\end{align}
with $\bm{D}_{\mu}=D\bm{e}^{z}\times\bm{e}^{\mu}$. To stabilize the linear spin-wave calculation, we introduce a small artificial field $\Delta h$ that favors a given spin configuration. This \textit{ad hoc} treatment is expected to be reasonable when the difference between the given spin configuration and the true ground state is very small but cannot be captured within the magnetic unit cell~\cite{takeda2024ncom}. We introduce the Fourier transformation of $\hat{b}_{\bm{r}}$ for $\bm{r}\in\Lambda_{\alpha}$ as
\begin{align}
    \hat{b}_{\bm{r}}=\frac{1}{\sqrt{N/N_{\mathrm{m}}}}\sum_{\bm{k}\in\mathrm{BZ}}\hat{b}_{\bm{k},\alpha}\mathrm{e}^{i\bm{k}\cdot\bm{r}},
\end{align}
and up to the constant term, the Hamiltonian~(\ref{eq:Hmag_real}) is reduced to
\begin{align}
    \hat{\mathcal{H}}_{\mathrm{mag}}=\frac{1}{2}\sum_{\bm{k}}\hat{\Phi}_{\bm{k}}^{\dagger}H_{\mathrm{BdG}}(\bm{k})\hat{\Phi}_{\bm{k}},
\end{align}
where $\hat{\Phi}_{\bm{k}}=(\hat{b}_{\bm{k},1},\cdots,\hat{b}_{\bm{k},N_{m}},\hat{b}_{-\bm{k},1}^{\dagger},\cdots,\hat{b}_{-\bm{k},N_{m}}^{\dagger})^{T}$, and $H_{\mathrm{BdG}}(\bm{k})$ is the bosonic Bogoliubov de-Gennes (BdG) Hamiltonian
\begin{align}
    H_{\mathrm{BdG}}(\bm{k}) =
    \begin{pmatrix}
        \Xi(\bm{k}) & \Delta(\bm{k})\\
        \Delta^{*}(-\bm{k}) & \Xi^{*}(-\bm{k})
    \end{pmatrix},
\end{align}
whose matrix elements are calculated as ($k_{\mu}=\bm{k}\cdot\bm{e}_{\mu}$)
\begin{align}
    \Xi_{\alpha,\beta}(\bm{k}) &= \varepsilon_{\alpha}\delta_{\alpha,\beta} + \sum_{\mu}\left(t_{\alpha,\beta}\mathrm{e}^{ik_{\mu}}+t_{\beta,\alpha}^{*}\mathrm{e}^{-ik_{\mu}}\right)\delta_{\alpha_{\mu},\beta}, \\
    \Delta_{\alpha,\beta}(\bm{k}) &= \sum_{\mu}\left(\Delta_{\alpha,\beta}\mathrm{e}^{ik_{\mu}}+\Delta_{\beta,\alpha}\mathrm{e}^{-ik_{\mu}}\right)\delta_{\alpha_{\mu},\beta}.
\end{align}
The bosonic BdG Hamiltonian $H_{\mathrm{BdG}}(\bm{k})$ is Hermitian since $\Xi(\bm{k})$ and $\Delta(\bm{k})$ satisfy $\Xi^{\dagger}(\bm{k})=\Xi(\bm{k})$ and $\Delta^{T}(\bm{k})=\Delta(-\bm{k})$. To diagonalize the Hamiltonian, we perform the Bogoliubov transformation
\begin{align}
    \hat{\Phi}_{\bm{k}} = P(\bm{k})\hat{\Psi}_{\bm{k}},
\end{align}
where $\hat{\Psi}_{\bm{k}}=(\hat{\beta}_{\bm{k},1},\cdots,\hat{\beta}_{\bm{k},N_{m}},\hat{\beta}_{-\bm{k},1}^{\dagger},\cdots,\hat{\beta}_{-\bm{k},N_{m}}^{\dagger})^{T}$ is the set of new bosonic operator and $P(\bm{k})$ is the paraunitary matrix satisfying
\begin{align}
    P^{\dagger}(\bm{k})\Sigma^{z}P(\bm{k})=P(\bm{k})\Sigma^{z}P^{\dagger}(\bm{k})=\Sigma^{z},
\end{align}
with $\Sigma^{\mu}=\tau^{\mu}\otimes I_{N_{\mathrm{m}\times N_{\mathrm{m}}}}$ ($\mu=x,y,z$). From the solution of the eigenvalue equation
\begin{align}
    \Sigma^{z}H_{\mathrm{BdG}}(\bm{k})\bm{p}_{n}(\bm{k}) = \varepsilon_{n}(\bm{k})\bm{p}_{n}(\bm{k}),
\end{align}
the paraunitary matrix can be constructed as
\begin{align}
    P(\bm{k}) = \left(\bm{p}_{1}(\bm{k}),\cdots,\bm{p}_{N_{m}}(\bm{k}),\Sigma^{x}\bm{p}_{1}^{\dagger}(-\bm{k}),\cdots,\Sigma^{x}\bm{p}_{N_{m}}^{\dagger}(-\bm{k})\right).
\end{align}
Then the Hamiltonian is reduced to the following form
\begin{align}
    \hat{\mathcal{H}}_{\mathrm{mag}}=\frac{1}{2}\sum_{\bm{k}}\sum_{n=1}^{N_{\mathrm{m}}}\varepsilon_{n}(\bm{k})\left(\hat{\beta}_{\bm{k},n}^{\dagger}\hat{\beta}_{\bm{k},n}+\frac{1}{2}\right).
\end{align}

\end{document}